\documentclass[12pt,preprint]{aastex}

\newcommand{\om}{\ensuremath{\Omega_m}}

\newcommand{\ola}{\ensuremath{\Omega_{\Lambda}}}
\newcommand{\kmsmpc}{{\ensuremath{{\rm km~s}^{-1}~{\rm Mpc}^{-1}}}}

\newcommand{\cosmol}[3]{\ensuremath{\om = #1, \ola = #2, H_0 = #3~\kmsmpc}}
\newcommand{\lcdmparm}{\cosmol{0.27}{0.73}{71}}
\newcommand{\etal}{et~al.\/}

\newcommand{\msun}{\ensuremath{M_\sun}}
\newcommand{\msunpyr}{M\ensuremath{_\odot}\,yr\ensuremath{^{-1}}}

\newcommand{\ie}{i.\,e.}

\newcommand{\egc}{e.\,g.,}
\newcommand{\citeeg}[1]{\citep[\egc][]{#1}}

\newcommand{\rmn}[1]{\uppercase\expandafter{\romannumeral #1}}

\newcommand{\lstar}{L\ensuremath{_\star}}
\newcommand{\wphz}{W\,Hz\ensuremath{^{-1}}}

\shorttitle{Radio AGN in SDSS clusters}
\shortauthors{Croft \etal}

\begin{document}

\title{Radio AGN in 13,240 galaxy clusters from the Sloan Digital Sky Survey}

\author{Steve Croft, Wim de Vries and Robert H.\ Becker}
\affil{Institute of Geophysics and Planetary Physics, Lawrence Livermore National Laboratory L-413, 7000 East Avenue, Livermore, CA 94550}
\affil{University of California, 1 Shields Avenue, Davis, CA}

\begin{abstract}
We correlate the positions of 13,240 Brightest Cluster Galaxies (BCGs) with $0.1 \le z \le 0.3$ from the maxBCG catalog with radio sources from the FIRST survey to study the sizes and distributions of radio AGN in galaxy clusters. We find that 19.7\%\ of our BCGs are associated with FIRST sources, and this fraction depends on the stellar mass of the BCG, and to a lesser extent on the richness of the parent cluster (in the sense of increasing radio loudness with increasing mass). The intrinsic size of the radio emission associated with the BCGs peaks at 55\,kpc, with a tail extending to 200\,kpc. The radio power of the extended sources places them on the divide between FR~I and FR~II type sources, while sources compact in the radio tend to be somewhat less radio-luminous. We also detect an excess of radio sources associated with the cluster, instead of with the BCG itself, extending out to $\sim 1.4$\,Mpc. 

\end{abstract}

\keywords{galaxies: active --- galaxies: clusters: general}

\section{Introduction}

Recent simulations suggest that the colors of massive galaxies in the local Universe can only be explained if AGN feedback quenches star formation in the host galaxy \citep{croton:06}. The link between supermassive black hole and galaxy formation \citep{hk:00} may depend critically on AGN feedback, via coupling of the mechanical power (via winds or jets) of AGN to the baryonic component of forming galaxies \citep{rj:04}, and radio jets are known to both quench \citep{springel:05} and enhance \citep{mo} star formation. Mergers, galaxy harassment, and other processes invoked in AGN triggering, are common in clusters.%

Studies of the AGN and galaxy populations, kinematics, and substructures in clusters provide insight into how galaxies, AGN and clusters of galaxies form and evolve.

Clusters also contain some of the most luminous galaxies in the local Universe, the Brightest Cluster Galaxies (BCGs), which are more likely to be radio-loud than other cluster galaxies \citep{vdl:07}. Feedback from the dissipation of the energy of expanding radio bubbles from such AGN has been proposed as a solution \citeeg{bk:02} to the cooling-flow problem (the lack of evidence for the extreme inflows of cold gas which are predicted to occur in clusters due to their short radiative cooling timescales).

\citet{best:07} used a sample of 625 nearby groups and clusters of galaxies selected from the SDSS \citep{sdss}. They showed that galaxies close to the centers of clusters have an enhanced probability of being radio-loud AGN compared to field galaxies of similar mass. \citet{lm:07} studied a sample of 573 X-ray luminous clusters, and also found an enhancement of the fraction of galaxies that were radio-loud, as well as tentative evidence for an increase in this fraction with the mass of the cluster.

In this paper, we use a cluster sample more than 20 times larger than those used for the above studies. The maxBCG \citep{maxbcg} red sequence galaxy cluster finder was used by \citet{maxcat} to create a catalog of 13,823 galaxy clusters from the SDSS -- the largest galaxy cluster catalog assembled to date. The clusters have redshifts between 0.1 and 0.3, and contain between 10 and 190 red sequence galaxies brighter than 0.4\,\lstar\ within the virial radius.

We correlate the BCG positions with the FIRST \citep{first} radio source catalog, in order to study the BCGs which are radio-loud, as well as the spatial distribution of radio sources in the surrounding clusters. Of the 13,823 clusters in the SDSS sample, 13,240 are within the region covered by FIRST.

\label{sec:cosmo}We assume an \lcdmparm\ cosmology \citep{wmap}.

\section{Radio sources in the cluster environment}

\begin{figure}
\centering
\includegraphics[bb=31 170 600 420,clip=true,width=\linewidth]{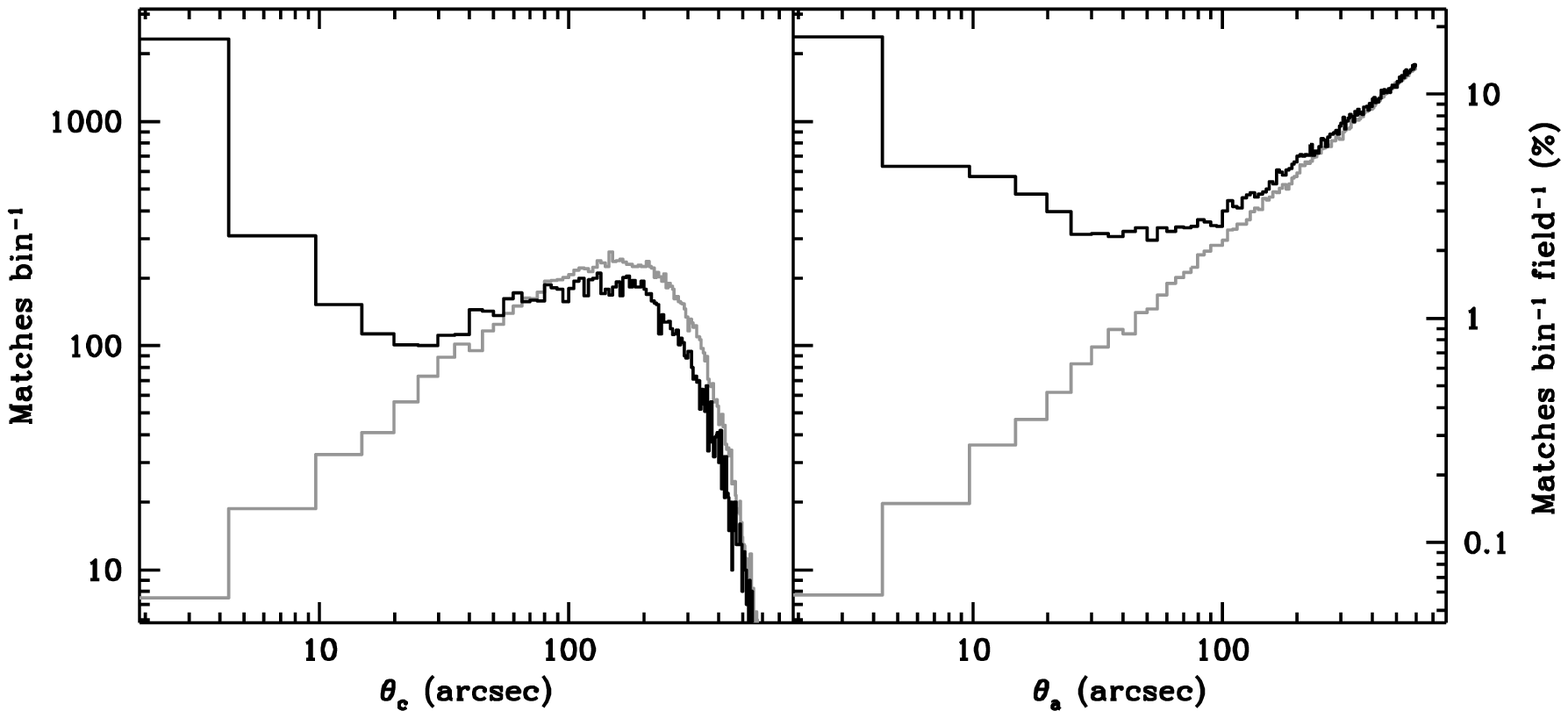}
\caption{\label{fig:excess}Histogram of the total number of FIRST sources in all 13,240 BCG fields, as a function of angular distance $\theta$ (in arcseconds) from the target (BCG or offset comparison field) position, in bins (\ie, annuli) of width 5\arcsec. The black histogram shows the counts in the fields containing the BCG, and the grey histogram shows the mean counts in the comparison fields. The left panel shows just the closest FIRST source to the target position (\ie, for each field, we consider only the single FIRST source closest to the target), whereas the right panel shows {\em all} FIRST sources in the target field (which can occasionally include multiple FIRST sources in a given annulus). The right-hand vertical axis in both panels shows the counts per bin as a percentage of the total number of fields (13,240). 
}
\end{figure}

We begin our analysis by finding the angular distance $\theta_c$ from the BCG position to the closest FIRST source. A histogram of the results for all 13,240 fields is shown in the left panel of Fig.~\ref{fig:excess}, along with a histogram of the mean number of random matches obtained by repeating this task around 4 positions offset by $\pm 30$\arcmin\ in the cardinal directions from each of the BCG positions. We also express the number of matches as a percentage of the total number of BCGs. Note that with increasing $\theta_c$, the BCG field counts initially fall below the background counts, due to the tendency of BCG fields to have a FIRST source closer to the BCG position than the average comparison field (which is as expected due to conservation of number). The background counts peak at $\theta_c = 147\farcs5$, which corresponds to the average distance from the comparison field position to the nearest FIRST source.

The first bin ($0 \leq \theta <  5$\arcsec) contains 2328 sources (compared to a mean of $7.5 \pm 1.7$ in the comparison fields); \ie, 17.5\%\ of fields contain a FIRST source within 5\arcsec\ of the BCG position, most of which are likely associated with an AGN hosted by the BCG. An additional 287 BCGs appear to host extended double-lobed radio sources (\S~\ref{sec:frii}). Therefore at least 19.7\%\ of BCGs in our sample are associated with FIRST sources.\label{sec:rlfrac}

Some of these extended sources show up as additional counts in the $\theta \lesssim 30$\arcsec\ bins of Fig.~\ref{fig:excess}. In the right-hand panel, we again count FIRST sources as a function of distance, but this time we count all matching FIRST sources, not just the closest (so that there are multiple FIRST matches per field, and sometimes more than one FIRST source per annulus). It is easy to show that in this case the number of matches per 5\arcsec\ bin for the comparison fields is given by $N = 10 \pi \rho r n$, where $\rho = 6.9 \times 10^{-6}$~sq.~arcsec$^{-1}$ is the surface density of FIRST sources, $r$ is the radius of the center of the annulus, and $n = 13240$ is the number of fields.

\begin{figure}
\includegraphics[width=\linewidth]{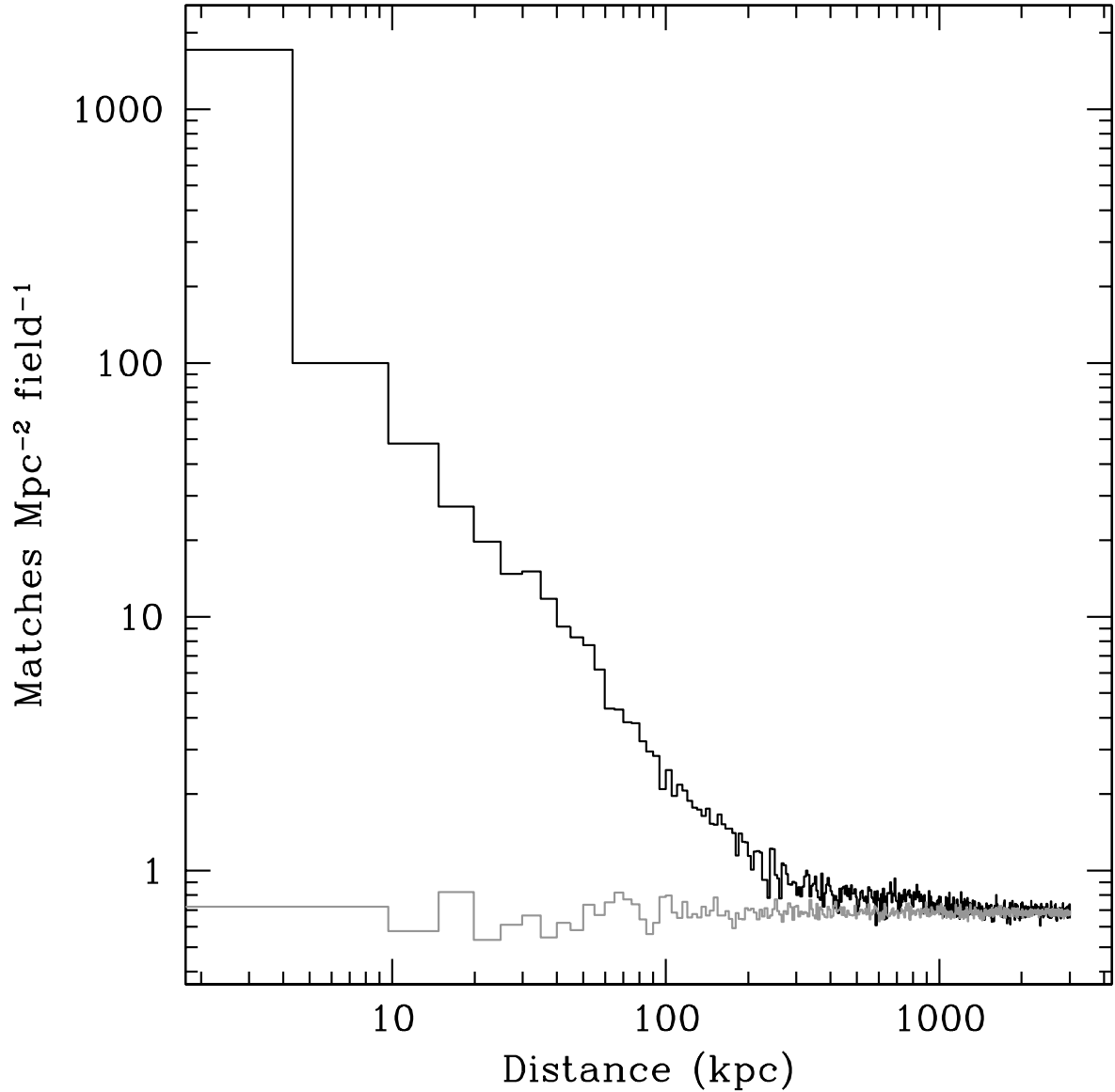}
\caption{\label{fig:realcounts}Histogram of the average projected surface density of FIRST sources (black), as a function of distance in kpc from the BCG position (assuming, for each field, that all FIRST sources are at the BCG photometric redshift). The grey histogram shows the comparison field counts, at the mean of four positions offset by 30\arcmin\ from the BCG positions. 
}
\end{figure}

At $\theta \gtrsim 30$\arcsec, the counts in the BCG fields begin to fall off less steeply than at smaller radii, as they cease to be dominated by the intrinsic size of the central source (see \S~\ref{sec:size}), and begin to be dominated by other radio galaxies within the cluster. At $\theta_a \gtrsim 400$\arcsec, the counts become indistinguishable from the background counts, suggesting that radio source activity is only enhanced (on average) within the central $\sim 1.4$\,Mpc of the cluster (at the mean BCG redshift for this sample $\bar{z} = 0.2184$; see also Fig.~\ref{fig:realcounts}).

The power of this analysis is that even with no assumptions about the redshifts of sources, or attempts to associate multiple components of individual radio galaxies with the ``parent'' optical source, one can see the excess of radio sources extends far out into the cluster. By contrasting the left and right panels of Fig.~\ref{fig:excess}, one can also see that there is a significant excess of multiple-component radio sources on scales $\lesssim 50$\arcsec, before the slope in the right panel changes at about $100$\arcsec, where the excess is mainly due to cluster radio sources unassociated with the BCG.

Since the maxBCG catalog provides accurate photometric redshifts for all the BCGs, we can use these to convert $\theta$ into real projected distances (and radio flux densities into radio luminosities; \S~\ref{sec:frii}). In this analysis (Fig.~\ref{fig:realcounts}), we assume that each radio source in a particular field is at the same redshift as the BCG. As can be seen from Fig.~\ref{fig:excess} (right panel), this is not true at large angular separations $\theta$, where most of the FIRST sources are unassociated ``background'' sources. We attempt to compensate for this by scaling the background counts in the same manner (where we use the BCG redshift to scale the counts in all four associated comparison fields, and then compute the histogram of background counts for the entire sample). At smaller separations, most of the sources really are associated with the BCG, \ie, they are at the cluster redshift. It would of course be possible to match the other radio sources to galaxies from SDSS and attempt to select only sources at similar photometric redshifts to the BCG. In this analysis we wish to concentrate on the ``pure'' radio population in the clusters, however, so we defer such radio--optical matching to future work.

In Fig.~\ref{fig:realcounts}, we show the average number of FIRST sources per field, divided by the area of each annulus, \ie, the surface density of radio sources in the clusters. Again, there is a prominent peak associated with the radio-detected BCGs, which has a wide shoulder extending to $\sim 200$\,kpc due to the presence of extended radio sources (see \S~\ref{sec:frii}). Even out to large radii, there is an excess of sources in the cluster environments compared to the comparison fields. Since we have such a large cluster sample, we are also able to bin by other parameters such as radio luminosity, redshift, cluster richness, etc., and look for changes in the distribution of radio sources. These results will be discussed in an upcoming paper.

\begin{figure}
\includegraphics[width=\linewidth]{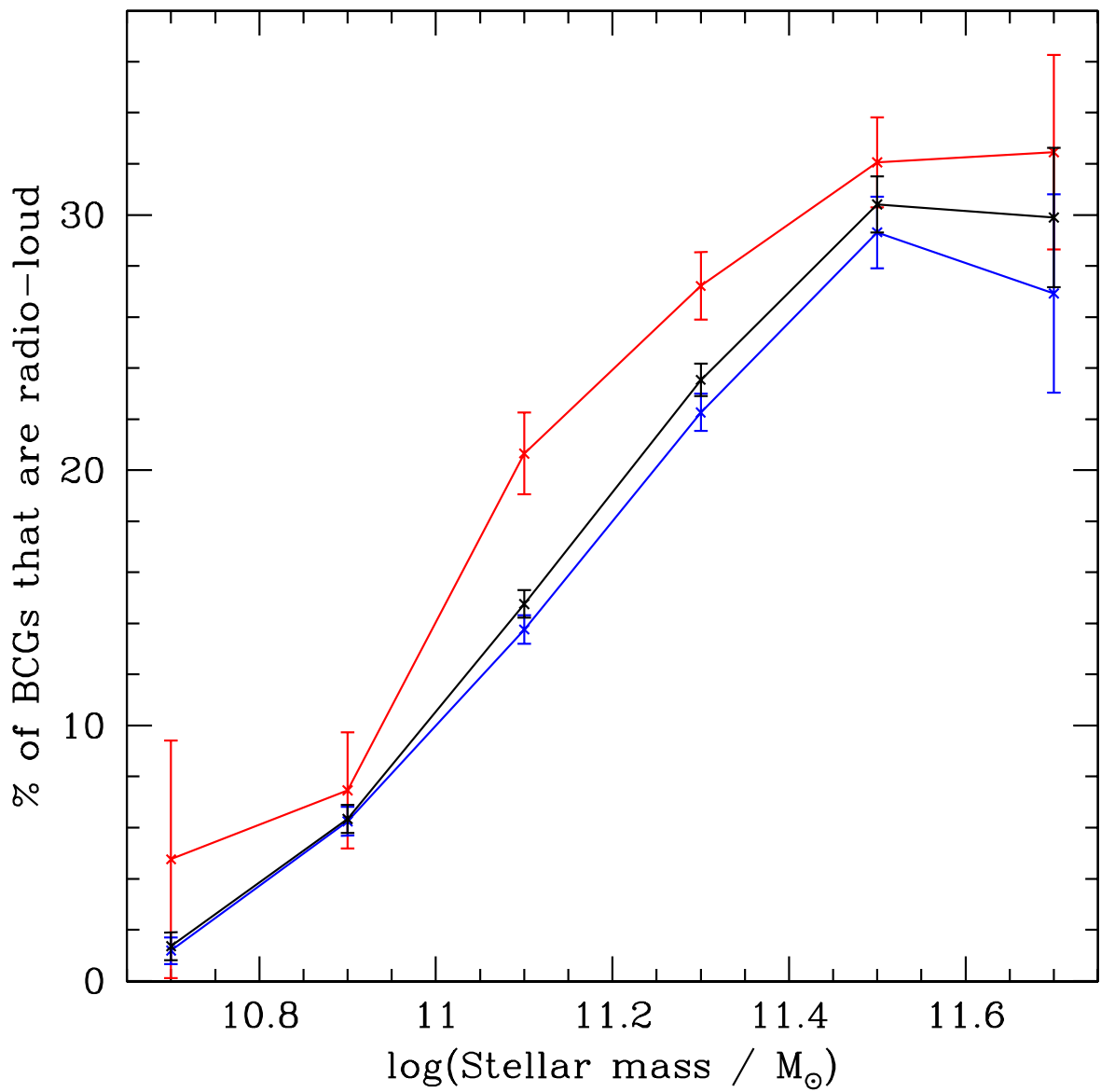}
\caption{\label{fig:rlfrac}The percentage of BCGs that are radio-loud AGN, as a function of stellar mass (black line). Also plotted are the same data split into two subsamples; BCGs in poor clusters ($< 20$ members; blue line) and rich clusters ($\geq 20$ members; red line). At a given stellar mass, BCGs in rich clusters have a slightly higher probability of being radio-loud.
}
\end{figure}

In Fig.~\ref{fig:rlfrac} we show the fraction of BCGs which are radio-loud (which, for consistency with \citealt{best:07}, we define as having an associated FIRST source with $L_{1.4} > 10^{23}$\,\wphz). \citet{maxcat} provide BCG luminosities in SDSS $r$-band, $k$-corrected to $z = 0.25$. We convert these luminosities to absolute magnitudes, $M_r$, and assume an $r$-band stellar mass-to-light ratio log\,$(M/L) = 0.5$ \citep[][Fig.~14]{kauffmann:03}. We recover the result from \citet{best:07} that the radio-loud fraction increases with increasing stellar mass. \citeauthor{best:07} found that BCGs in rich clusters were more likely to be radio-loud than those in poor clusters, but concluded that this was due to the tendency of the richest clusters to host the most massive BCGs. Using our larger sample, we are able to show that {\em for a given stellar mass} the probability of a BCG to be radio loud is enhanced somewhat in richer clusters, suggesting that the wider environment plays a role in radio-loudness, in agreement with the findings of \citet{lm:07}.

\section{BCG radio luminosities and morphologies}\label{sec:frii}

By converting the measured radio fluxes for the BCGs into rest-frame 1.4\,GHz radio luminosities ($L_{1.4}$) we can investigate the dependence of radio luminosity on the optical luminosity of the BCG. For consistency with previous literature \citeeg{condon:92,lm:07} we adopt an average radio spectral index $\alpha = -0.8$ ($S_\nu \propto \nu^{\alpha}$) when applying the $k$-correction to calculate $L_{1.4}$. At the relatively low redshifts considered in this paper, the choice of spectral index does not strongly affect our results.

\begin{figure}
\includegraphics[width=0.9\linewidth]{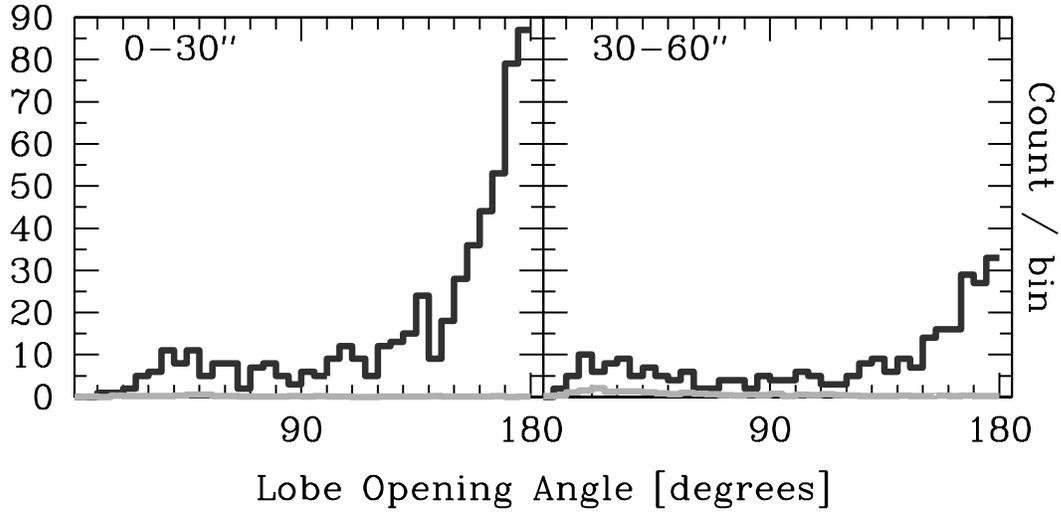}
\caption{\label{fig:frii}Histogram of opening angles for double and triple radio sources associated with BCGs (black) and expected contamination from random superpositions of sources which appear to be doubles / triples (grey). The left panel shows doubles / triples with major axes between 0 and 30\arcsec, and the right panel, major axes between 30 and 60\arcsec. The bin size is 5\degr.
}
\end{figure}

In order to account for extended emission, we use the technique of \citet{frii} to find double-lobed radio sources associated with the BCGs, even where the BCG does not have a radio core detected by FIRST. In Fig.~\ref{fig:frii}, we plot the lobe opening angle ($\Psi$, the angle subtended at the BCG position by the FIRST sources identified as lobes; see \citeauthor{frii} for more details) for double-lobed radio sources in two size bins. In the subsequent discussion, we restrict our analysis to double-lobed sources with $\Psi >100$\degr, where the fraction of false positives (shown as the blue histogram in Fig.~\ref{fig:frii}) is low. At small angles ($\sim 45$\arcdeg) we also see a small secondary peak in the cluster fields, due an excess (compared to the comparison fields) of real double sources in the cluster which are associated with galaxies other than the BCG. Due to their true hosts being offset from the BCG position, these are sometimes erroneously identified by the algorithm as being small opening-angle sources associated with the BCG itself (but again we emphasise that most sources with $\Psi > 100$\degr\ are real). We find 453 BCGs (compared to a background expectation of 1.00) associated with double-lobed radio sources with $\Psi > 100$\degr\ and diameter $0 - 30$\arcsec\ (83 where the core is also detected, and 370 where it is not). Additionally, we find 196 BCGs (compared to 4.90 in the background case) with double-lobed radio sources with $\Psi > 100$\degr\ and diameter $30 - 60$\arcsec\ (108 with cores, and 88 without). Note that 362 of the 649 BCGs associated with doubles were also counted in the 2328 sources with a FIRST source within 5\arcsec\ discussed in \S~\ref{sec:rlfrac}. 

\begin{figure}
\includegraphics[width=0.9\linewidth]{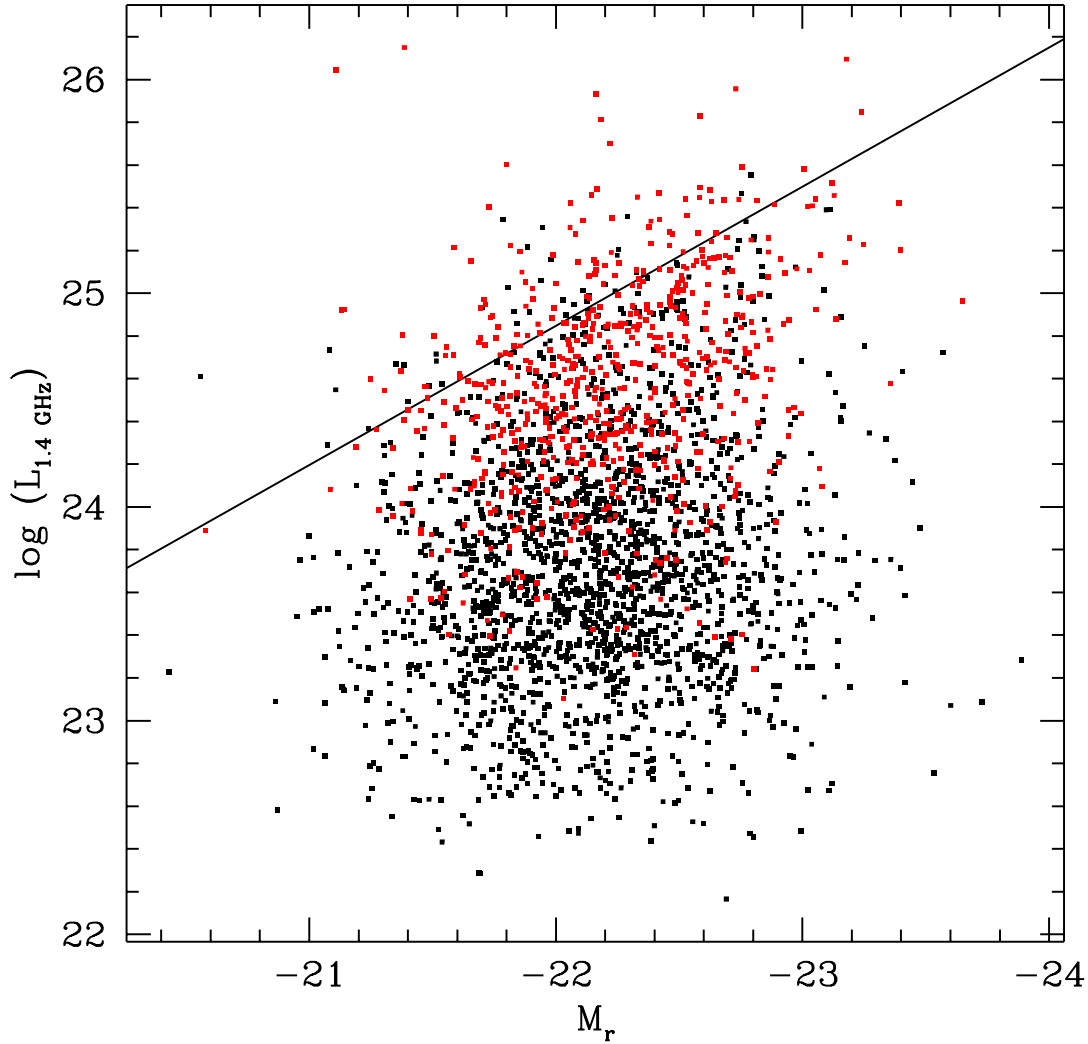}
\caption{\label{fig:lo}Rest-frame 1.4\,GHz radio luminosity as a function of SDSS $r$-band absolute magnitude for the BCGs detected in FIRST, $k$-corrected to $z = 0.25$. BCGs with a FIRST source within 5\arcsec\ are shown as black squares, except for cases where our ``double-detector'' algorithm (see \S~\ref{sec:frii}) finds a double radio source associated with the BCG (sometimes with a radio-detected core, sometimes without). These 649 doubles and triples are plotted as red squares, with the radio luminosity determined from the total integrated radio flux of the components. The solid line is the division between FR~I and FR~II radio galaxies from \citet{lo:96} (we make no correction from their $M_{24.5}$ to our $M_r$).
}
\end{figure}

In Fig.~\ref{fig:lo}, we plot $L_{1.4}$ (inferred from the total fluxes of the radio components associated with each BCG) for the 649 double-lobed sources (red squares) as a function of $M_r$. The black squares in Fig.~\ref{fig:lo} are the 1966 BCGs {\em not} identified as doubles, which nevertheless have a FIRST source within 5\arcsec. We also plot the empirical division between FR~I and FR~II radio galaxies from \citet{lo:96}, where we have assumed $M_r = M_{24.5}$ (the isophotal magnitude to 24.5 mag arcsec$^{-2}$ from \citeauthor{lo:96}).

\begin{figure}
\includegraphics[width=0.9\linewidth]{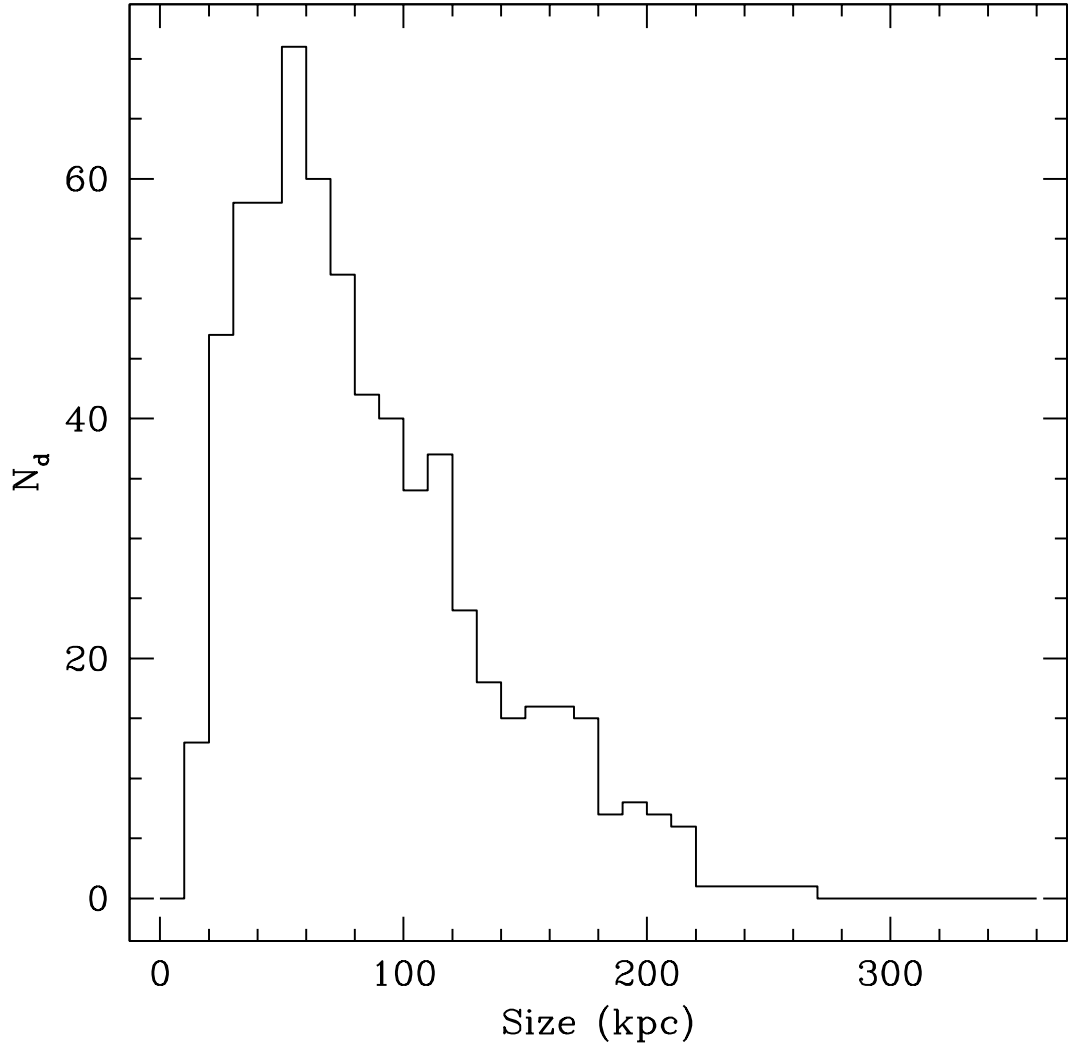}
\caption{\label{fig:friihist}$N_d$, the number of double-lobed radio sources as a function of physical size of the radio emission for the 649 doubles in our sample.
}
\end{figure}

In contrast to the results of \citet{frii} for SDSS quasars, we find very little excess at $\Psi \sim 180$\degr\ for diameters $60 - 120$\arcsec\ (\ie, there are very few doubles as large as this in our BCG sample). Since \citet{frii} studied quasars with median redshift $z = 1.40$ and associated radio emission from FIRST, their sample has a higher mean radio luminosity, so this result is perhaps unsurprising, even given the smaller angular diameter distance to our sources; the physical sizes of our sources are about a factor 4 smaller than the FR~IIs of \citeauthor{frii}, peaking at 55\,kpc (Fig.~\ref{fig:friihist}), \label{sec:size} and dropping to zero at around 200\,kpc. As can be seen from Fig.~\ref{fig:realcounts}, 200\,kpc is where the slope of the histogram changes, showing that the excess matches at $> 200$\,kpc are dominated by other radio galaxies in the cluster, and at $< 200$\,kpc, by the presence of multiple-component radio sources associated with the BCG.

Fig.~\ref{fig:lo} shows that most of the double-lobed sources are located near the FR~I~/~II division from \citet{lo:96}, which adds weight to the argument that radio AGN in cluster environments at low redshift do not tend to be powerful FR~IIs \citep{hl:91} -- they could certainly be seen by us if present. It is clear that the upper envelope in radio luminosity in this plot shows an increase in radio luminosity with optical luminosity (a proxy for stellar mass), in agreement with previous results \citeeg{jarvis:01}.

We note that only a few detected sources (7.2\%) are fainter than $10^{23}$\,\wphz\ (essentially due to the FIRST flux limit). For such a radio luminosity to be due entirely to star formation would require SFR $\sim 100$\,\msunpyr\ \citep{hodge:07}, so we conclude that radio-loud AGN are the dominant source of radio emission in those of our BCGs detected in FIRST. Of course, even in sources with $S_{1.4} \gg 1$\,mJy, star formation may contribute to the radio emission. In some cases, this star formation will also manifest as optical emission lines, whereas in other cases the star formation may be optically obscured.

\section{Summary and future work}

By matching the maxBCG and FIRST catalogs, we were able to determine that 19.7\%\ of BCGs in our sample are detected in the radio; the majority (92.8\%) of these are radio-loud AGN ($L_{1.4} \geq 10^{23}$\,\wphz). The radio loud fraction has a strong dependence on stellar mass of the host, increasing from $\sim 1$\%\ at $4 \times 10^{10}$\,\msun\ to $\sim 30$\%\ at $5 \times 10^{11}$\,\msun. For a given stellar mass, BCGs in rich clusters have a higher tendency to be radio-loud than BCGs in poor clusters or groups, suggesting that the wider environment has some influence on radio-loudness.

Signs of double-lobed emission are seen (at the resolution of FIRST) in 24.8\%\ of the 2615 BCGs associated with FIRST sources. The physical size of these double-lobed sources peaks at 55\,kpc, with a tail extending to $\sim 200$\,kpc. Their radio luminosities place them close to the FR~I~/~II division.

There is an excess of radio sources in our clusters, which are associated with other cluster members, extending out to $\sim 1.4$\,Mpc, but the central BCGs are, overall, the dominant source of AGN activity in these clusters.

Future investigations using this dataset could address the relationship between optical spectroscopic indicators of star formation and AGN activity (as determined from the SDSS ``value added'' catalogs of \citealt{brinchmann:04} and \citealt{agncat}), radio loudness / luminosity, and environment, as well as extending the FIRST-SDSS match to other galaxies in the clusters (as determined by their photometric redshifts).

For those BCGs undetected in FIRST, it is possible to ``stack'' the FIRST maps \citep{stacking,stacking2} to determine the average radio properties of sources too faint to be detected individually. In this way, we can study BCGs with radio emission two orders of magnitude below the FIRST flux limit. These results will be discussed in a future paper.

\acknowledgments

Work was performed under the auspices of the U.\ S.\ Department of Energy, National Nuclear Security Administration by the University of California, Lawrence Livermore National Laboratory under contract No.\ W-7405-Eng-48. Research made use of the VLA FIRST Survey (NRAO programs AB628, AB879 and AB950) and the SDSS. The National Radio Astronomy Observatory is a facility of the National Science Foundation operated under cooperative agreement by Associated Universities, Inc.

\clearpage

\clearpage

\clearpage

\clearpage

\clearpage

\clearpage

\end{document}